\documentclass[conference]{IEEEtran}
\usepackage{cite}
\usepackage{amsmath,amssymb,amsfonts}
\usepackage{graphicx}
\usepackage{textcomp}
\usepackage{siunitx}
\usepackage{algorithm}
\usepackage{algpseudocode}
\usepackage{fancyhdr}
\usepackage{xcolor}
\def\BibTeX{{\rm B\kern-.05em{\sc i\kern-.025em b}\kern-.08em
    T\kern-.1667em\lower.7ex\hbox{E}\kern-.125emX}}

\fancypagestyle{cfooter}{ %
\fancyhf{} 
\cfoot{\scriptsize{This work has been submitted to the IEEE for possible publication. Copyright may be transferred without notice, after which this version may no longer be accessible.}
}

}

\begin{document}

\title{Unsupervised high-throughput segmentation of cells and cell nuclei in quantitative phase images}

\author{\IEEEauthorblockN{Julia Sistermanns\IEEEauthorrefmark{1} \, \, \, Ellen Emken\IEEEauthorrefmark{2} \, \, \, Oliver Hayden\IEEEauthorrefmark{2} \, \, \, Gregor Weirich\IEEEauthorrefmark{3} \, \, \, Wolfgang Utschick\IEEEauthorrefmark{1} \\\\}
\IEEEauthorblockA{\IEEEauthorrefmark{1}\textit{Chair of Methods of Signal Processing, Department of Computer Engineering, TUM}, Munich Germany \\
\IEEEauthorrefmark{2}\textit{Heinz-Nixdorf-Chair of Biomedical Electronics, Department of Electrical Engineering, TUM}, Munich Germany \\
\IEEEauthorrefmark{3}\textit{Institute of Pathology, School of Medicine and Health, Technical University of Munich}, Munich Germany \\
Email Corresponding Author: julia.sistermanns@tum.de}
}

\maketitle
\thispagestyle{cfooter}

\begin{abstract}
In the effort to aid cytologic diagnostics by establishing automatic single-cell screening using high-throughput digital holographic microscopy for clinical studies thousands of images and millions of cells are captured. The bottleneck lies in an automatic, fast, and unsupervised segmentation technique that does not limit the types of cells which might occur. We propose an unsupervised multistage method that segments correctly without confusing noise or reflections with cells and without missing cells that also includes the detection of relevant inner structures, especially the cell nucleus in the unstained cell. In an effort to make the information reasonable and interpretable for cytopathologists, we also introduce new cytoplasmic and nuclear features of potential help for cytologic diagnoses which exploit the quantitative phase information inherent to the measurement scheme. We show that the segmentation provides consistently good results over many experiments on patient samples in a reasonable per-cell analysis time.
\end{abstract}


\section{Introduction}
The success of cervical smear screening (PAP-test) has proven that cytopathology is capable of detecting cancer precursors and malignant changes in the very early stages of a disease\cite{DeMay:1996ux,Takahashi:2000tg}. The gold standard of the cytologic diagnostic workflow relies on manual sample smearing, methanol fixation and cell labelling by the standardized PAP stains, and coverslipping before changes can be looked for under a microscope by a trained cytopathologist~\cite{Al-Abbadi:2011tl}. A single cytopathologist will check several hundred cells, but cytological samples easily contain thousands of cells, and especially in the early stages of disease only very few cells are affected~\cite{DeMay:1996ux,Takahashi:2000tg}. 

A lot of effort is going into research to aid cytologic diagnostics with new imaging and measurement schemes to automatically collect information-rich measurements of cells followed by automatic single-cell analysis. Instead of looking at just a selection of cells, all cells of a sample could be checked. 
To establish automated cytologic screening, large clinical studies with more than 100 patients are necessary. This need is challenged by the fact that a single sample will provide thousands of images. The sheer data abundance present renders manual information processing infeasible~\cite{Meijering:2012aa}. An efficient, unsupervised cell segmentation technique is the key to overcome this bottleneck dilemma by extracting the actual single-cell information from these images while discarding noise and measurement artifacts.\par 
Cell segmentation is not a new challenge, quite the contrary, research has been ongoing since the 1960s and a variety of algorithms exist and are being continuously improved upon~\cite{Meijering:2012aa, Loewke:2018aa, Vicar:2019aa}. The fact that still to this day research is ongoing and human-aided interactive approaches~\cite{Alemi-Koohbanani:2020aa,Korzynska:2007aa,Papadopoulos:2017aa,Wang:2018aa,Amgad:2022aa} are being proposed, is a testament that no one correct solution could be found so far. The main automated cell segmentation approaches can be grouped into the following classes: intensity thresholding~\cite{Awasthi:1994aa}, feature detection~\cite{Meijering:2012aa}, mathematical morphology filtering~\cite{Anoraganingrum:1999aa,Beucher:1979aa,Gowda:2017aa,Shen:2018aa}, region growing~\cite{Adams:1994aa,Ram:2012aa,Wittenberg:2004aa}, level-set or active contour methods~\cite{Bar:2011aa,Chan:2001aa,Chan:2006aa,Chung:2009aa,Ersoy:2008aa,Osher:2003aa,Rada:2013aa,Tsai:2005aa} and using deep learning approaches trained on ground truth labeled datasets~\cite{Gao:2023aa,Lee:2020aa,Schmidt:2018sd,Song:2017aa,Vicar:2021ss}. For quantitative phase microscopy (QPM), which is the imaging technique applied in this work, several specialized methods have been proposed: an iterative thresholding technique matching known volumetric distributions~\cite{Loewke:2018aa} which was further improved by additional Laplacian of Gaussian image enhancement and distance transform-based splitting~\cite{Vicar:2021aa}, and a deep-learning-based approach for detecting platelet aggregates trained on simulated data~\cite{Klenk:2023ab}. All the proposed approaches rely on the fact that the occurring cells or cell aggregates have an appearance that is, to some degree, expectable and a ground truth for training or calibrating the algorithm can be found. Also, they focus solely on detecting the cells and not on detecting the nucleus of the cells.


Our contribution is an unsupervised multistage segmentation for quantitative phase microscopy which is automatic, fast, and unsupervised and includes the detection of relevant inner structures, especially the cell nucleus in the unstained cell. Additionally, we introduce new morphological features that exploit not only the structural information of the cell but also the optical density information gained from the use of QPM and mimic morphologic characteristics cytopathologists search for in their diagnosis.


\section{Data}
\label{sec:data}

\subsection{Cytologic Samples and Diagnostic Process}

Most cytologic samples harbour different cell types in distinctive life cycle stages with repercussions on morphology. Adding to the plethora of benign variations, and their differential diagnosis to cancer, cancer  by itself is not a single disease. Accordingly, there is no single cancer qualifier of the gestalt, of the proteome, or of the genome~\cite{DeMay:1996ux}. Most cancers evolve by sequential gene mutations, many of them leaving morphologic fingerprints which can be detected by microscopy of single cells, i.e. cytopathology, in nuclear and cytoplasmic details~\cite{Nauth:2014gy, Mody:2018wi, Takahashi:2000tg}. However, as precursor lesions are rare, they may evade microscopic detection, such that new screening methods capable of registering the entirety of a cellular sample are in demand.

We will show quantitative phase images of cervical smears. The samples are patient samples taken from an ongoing clinical study in cooperation with the Institute of Pathology and Department of Gynecology and Obstetrics of the Technical University of Munich at the University Hospital Rechts der Isar in parallel to standard diagnostics. At the time of segmentation, the cell types and stages of disease present in the samples are unknown, vary greatly from sample to sample and the cells are still viable. 

\subsection{Optical Setup}
\label{sec:dhm}
The quantitative phase information of the cells presented in this work was captured with high-throughput digital holographic microscopy, which was also applied successfully in~\cite{Klenk:2023ab} to find a new biomarker in blood samples to predict COVID-19 severity and in~\cite{Ugele:2018ac} to detect leukemia. 
The liquid samples, in our study from cervical smears, are diluted in a solvent directly after sample acquisition and then introduced into a microfluidic channel. The sample flow is focused through hydrodynamic focusing to create a monolayer of cells. The customized digital holographic microscope has an off-axis configuration as described in~\cite{Klenk:2023ab}. It uses self-referencing and double-shearing techniques with a partially coherent light source with wavelength $\lambda = \SI{528}{\nm}$. The interference pattern is recorded by a digital camera as a so-called hologram. From this hologram, the amplitude and phase were reconstructed with the OsOne Software Version 5.12.12 of Ovizio Imaging Systems~\cite{Ovizio:51212}. 
The  local quantitative phase values $\phi (x,y)$ can be related directly to the so-called optical density $\rho_o(x,y)$~\cite{Girshovitz:2012aa}, which is a combination of the local thickness of the cell and difference in refractive indexes between the cell and microfluidic solution, through the equation
\begin{align}
    \phi (x,y) = \frac{2\pi}{\lambda}\rho_o(x,y)\,.
\end{align}
The optical density information of the cells provides valuable insights about the cells structure~\cite{Girshovitz:2012aa}.

\section{Segmentation Method}
\label{sec:method}

\begin{figure*}[htbp]
\centerline{\includegraphics[scale=0.5]{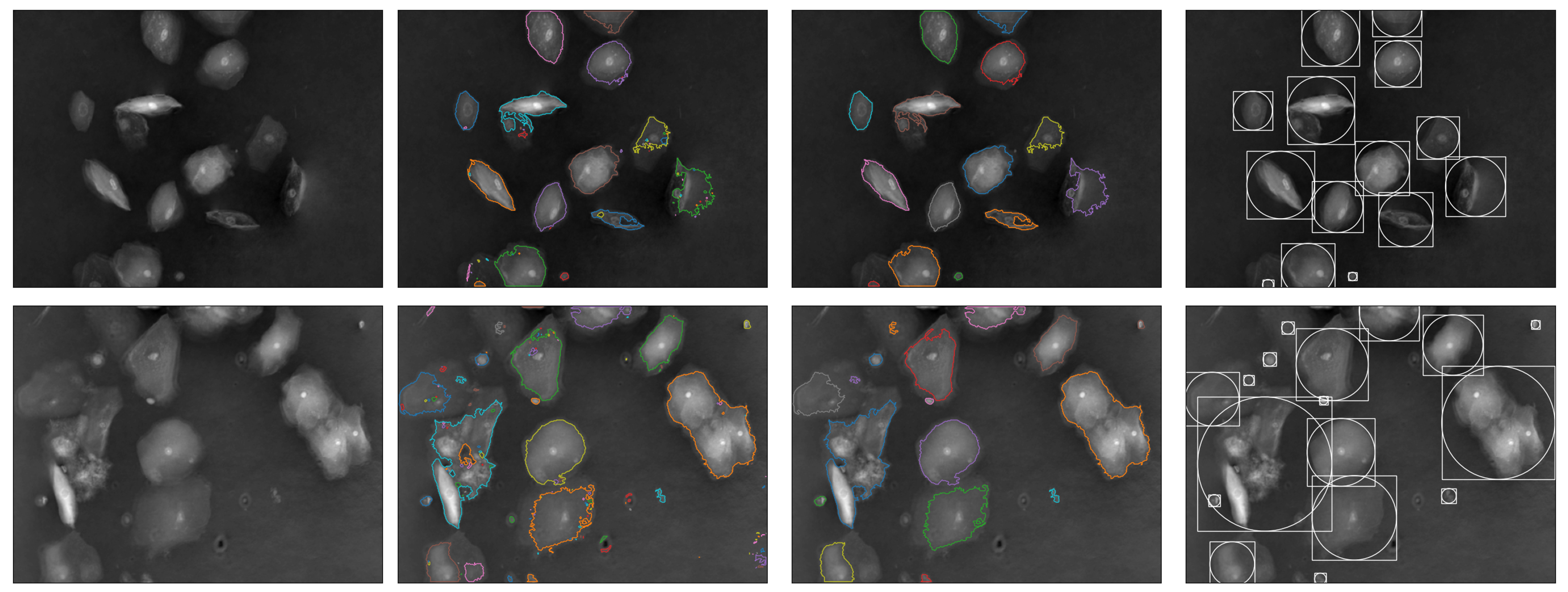}}
\caption{Two Examples of the proposed cell detection process (section~\ref{sec:statistics}-~\ref{sec:check}). From left to right: (1) quantitative phase image, (2) image after the threshold segmentation (step 1$\&$2), (3) plausibility check (step 3) which discards contours which are too small, lie inside another contour or are noise or debris; (4) To better highlight the detected cells the enscribing circle around the contours and the bounding box are highlighted.}
\label{fig}
\end{figure*}

The proposed segmentation method follows a multistage approach to save computation time while continuously refining the found contours and is shown for two examples in Fig.~\ref{fig}. The first step is a statistical analysis of the absolute phase values of all captured images of a single measurement, which is then used for an automatic choice of an appropriate threshold value for step two, which is a simple threshold segmentation.
In step three, all found contours are checked for plausibility under certain constraints of the measurement set-up. After step three, the internal structures are detected inside each individual cell in step four, and a variety of morphological, structural, textural, nuclear, and cytoplasmic features are extracted.

\subsection{Statistical Analysis of Phase Values}
\label{sec:statistics}
The statistical analysis of the phase values of all images of a measurement is the most time-consuming step, but it makes automatic decisions for the detection and filtering of artifacts (like reflections and phase wraps) possible. Only 4 values extracted from each individual image $i$ are necessary: the minimal, maximal and mean phase value of the image $\phi_{i\text{, min}}$, $\phi_{i\text{, max}}$ and $\bar{\phi}_{i}$ and the background color of the image $c_{b,i}$. The background color of each image is chosen to be the phase value which occurs most often in this image. Due to the measurement set-up, most images either have only a few cells or are empty, only very few exceptions occur where cells are covering the entire field of view. Given the large amount of images ($N \approx 60000$ per sample), this is a valid assumption.

\subsection{Threshold Stage}
\label{sec:thresh}
Step two, the detection of the cells in all images not filtered in step one is performed by a simple threshold segmentation, as this is by far the fastest option~\cite{Vicar:2019aa}. This step is intentionally kept coarse, so as to catch all possible candidates and not miss any cells. As the measurement set-up with the micro-fluidic channel provides a scenario with predominately well-separated cells, which are distinguishable from the background, such a simple segmentation is sufficient. The threshold is chosen automatically based on the previous statistical analysis as: 
\begin{equation}
    t = 2 |\bar{c}_b| \, \text{, with } \bar{c}_b = \sum_{i=1}^{N} c_{b,i} \, .
\end{equation} 

\subsection{Checking Mechanisms}
\label{sec:check}
In the next step, the detected contours are checked for plausibility with three constraints to identify and discard all non-cells (Fig.~\ref{fig}). The first constraint is that the contour area must be larger or equal to 
\begin{equation}
    A_{\text{min}} = \pi r_{\text{min}}^2
\end{equation}
where $r_{\text{min}} = \SI{3}{\um}$ for our study. This value corresponds to the lower bound of human erythrocytes with a diameter of $6$--$\SI{8}{\um}$~\cite{Ward:2018aa}. The cell diameter of squamous epithelial cells, which are the relevant cells for cervical cytology, is significantly larger~\cite{Hoda:2007pt, Nauth:2014gy, Mody:2018wi}.\\
\indent Secondly, cell contours which lie inside another detected contour are discarded. This occurs for example for cell aggregates or when a cell has a very irregular density distribution. In such a case the detection of internal structures will properly identify the internal structures.
The last check targets very noisy images, cell debris and slime, which cause extreme local differences of phase values and lead to false-positives of the threshold segmentation. A second threshold segmentation is performed on the image gradient. If no contour can be detected at the position indicated by the prior segmentation step in the gradient, then the contour is discarded. 

\subsection{Detection of Internal Structures}
\label{sec:internal}
In cervical cytology, the most relevant morphological features reside within the nucleus. Cancer and its precursors (dysplasia) show nuclear atypia of different grades, often paralleled by cytoplasmatic immaturity~\cite{Nauth:2014gy}. This is also true for most other carcinomas and their precursors. 

To identify the relevant internal structures inside a cell and identify the cell nucleus especially, a second round of segmentation is performed inside the detected cells. The process follows much the same approach as the detection of the cells themselves presented in~\ref{sec:thresh}-~\ref{sec:check}, but is slightly more sophisticated. The threshold segmentation is now performed using not just one, but 3 different thresholds which are chosen relative to the cell detection threshold $t$ as $t_{i,1} = 4 t$, $t_{i,2} = 6 t$ and $t_{i,3}$ which is relative to the mean of the maximum phase values inside all detected cells $\bar{\phi}_{i, \text{max}}$ as $t_{i,3} = 0.8 \bar{\phi}_{i, \text{max}}$. The minimum size of the structure is chosen to be 20 pixels, as the resolution for smaller structures would not be sufficient. If only one contour can be found through this detection, this structure is chosen to be the nucleus. If more than one internal contour is found, possible nuclei are chosen to be those which were found with the strictest detection threshold $t_{i,3}$. If this still leaves more than one possible nucleus, the cell is flagged to be either abnormal or a possible aggregate of cells and the nucleus is chosen to be the structure with the highest mean optical density.

\begin{figure}[htbp]
\centerline{\includegraphics[scale=0.39]{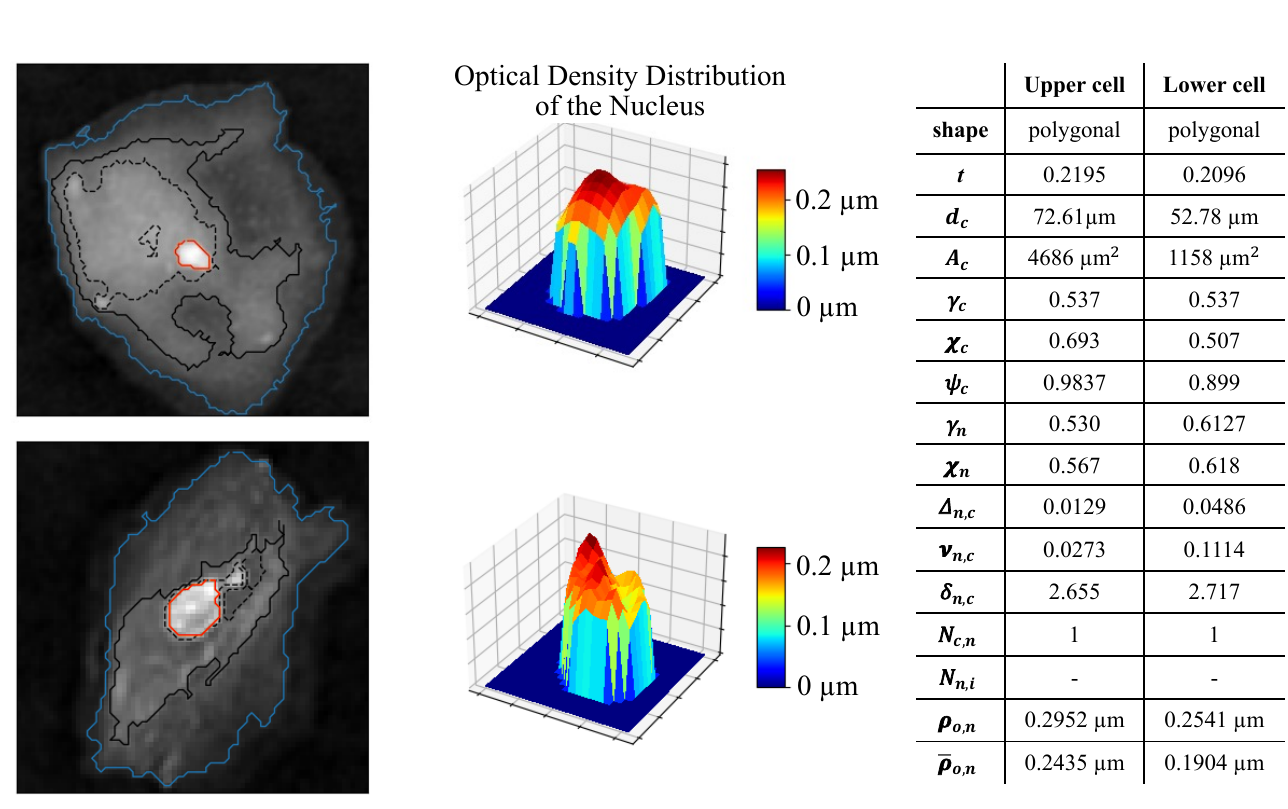}}
\caption{Two examples of the internal detection and feature extraction. Features: threshold $t$, cell diameter $d_c$, roundness $\gamma_c$, circularity $\chi_c$, polygonality $\psi_c$, roundness nucleus $\gamma_n$, circularity nucleus $\chi_n$, nuclear-cytoplasmic ratio $\Delta_{n,c}$, volumetric ratio $\nu_{n,c}$, position of nucleus $\delta_{n,c}$, number possible nuclei $N_{n,c}$, number internal structures in nucleus $N_{n,i}$, maximum and mean optical density nucleus  $\rho_{o,n}$ and $\bar{\rho}_{o,n}$}
\label{fig-examples}
\end{figure}

As this internal analysis is computationally expensive, the internal detection is only performed for cells which are most relevant to the diagnosis and have sufficient resolution. The minimum size was chosen as a diameter $d \geq \SI{25}{\um}$ based on standard values for epithelial cells from literature~\cite{Hoda:2007pt, Mody:2018wi, Nauth:2014gy}. Examples for the detection of internal structures, the choice of nucleus and the resulting new features are shown in Fig.~\ref{fig-examples} for two randomly chosen epithelial cells. 

\section{New Features}
\label{sec:features}
Based on the combination of the detected cell contour, the internal structures and the optical density information gained from the use of QPM, we introduce new morphological parameters that mimic morphologic characteristics cytopathologists search for in their diagnosis, pointing to cell classes and cytogenetic abnormalities. 

\subsection{Shape}
\label{sec:shape}
The cellular width and shape are essential for cell typing and cell maturity assessment~\cite {Hoda:2007pt, Mody:2018wi, Nauth:2014gy}. 
To detect the shape, for each cell a score of \textbf{circularity} $\chi_{c}$, \textbf{roundness} $\gamma_c$, \textbf{polygonality} $\psi_{c}$ and \textbf{ellipticity} $\varepsilon$ is calculated with values $\in [ 0, 1] $, where $0$ would mean that the cell doesn't fit this shape well and $1$ means that this shape fits the cell perfectly.

The shape scores are all found by comparing the actual area of the cell $A_c$ to the area of each shape $A_S$
\begin{align*}
\text{score} = \frac{\min\{A_c,A_S\}}{\max\{A_c,A_S\}}.
\end{align*}
To compute the areas of the shapes, the cells features are used, e.g. for the ellipticity, the cell is compared to the area of a perfect ellipsis where the axes correspond to the width and height of the cell.
\subsection{Nuclear and Cytoplasmic Features}
After the nucleus of the cell is detected, it is possible to extract several important features that pathologists search for in a cytologic diagnosis: the \textbf{nucleus diameter} $d_n$,  \textbf{nucleus area} $A_n$, the  \textbf{nucleus circularity} $\chi_n$ and  \textbf{roundness} $\gamma_n$. 
Nuclei of squamous epithelial cells are typically round or circular, nuclear unrounding may indicate dysplasia~\cite{Nauth:2014gy}. The next feature is the \textbf{position of the nucleus} $\delta_{n,c}$ in the cell which can be described through the Euclidean distance between the center point of the nucleus $(x_n, y_n)$ and the center point of the cell $(x_c, y_c)$.
If $\delta_{n,c}$ is small, the position is central, if $\delta_{n,c}$ is large, the nucleus will be close to the cell boundary and the position would be considered polar~\cite{Nauth:2014gy}. \\ Abnormal nuclear enlargement or shrinking for the type and stage of maturation of a cell, an important sign of dysplasia~\cite{Nauth:2014gy}, can be detected using the \textbf{nuclear-cytoplasmic size ratio} $\Delta_{n,c}$ comparing the nucleus and cytoplasm areas. Similarly, the volume of the nucleus $V_n$ can be compared to the volume of the cell $V_c$ as the \textbf{volumetric ratio}
\vspace{-0.1cm}
\begin{equation}
    \nu_{n,c} = \frac{V_n}{V_c}, \text{with } \,\, V_n = \sum\nolimits_{(x,y) \in n}  \rho_o (x,y) \cdot s^2_{\text{px}} \, .
\end{equation}
Another important diagnostic feature is the structural make-up of the nucleus: substructures, irregular density distributions of the nucleus, and a very high or low optical density in the nucleus, i.e., hyperchromasia or hypochromasia, can point to varying degrees of dysplasia~\cite{Nauth:2014gy}. Features to describe the structural make-up of the nucleus and quantify the optical density distribution of the nucleus are the \textbf{number of internal structures inside the nucleus} $N_{n,i}$, the \textbf{maximum and mean optical density of the nucleus} $\rho_{o,n}$ and $\bar{\rho}_{o,n}$.
Multinucleation, a typical response to inflammation or radiation~\cite{DeMay:1996ux}, is immediately detected by this internal segmentation and quantified as $N_{c, n}$ the \textbf{number of possible nuclei in a cell}.

\section{Results}
\label{sec:results}
\begin{figure}[tp]
\centerline{\includegraphics[scale=0.6]{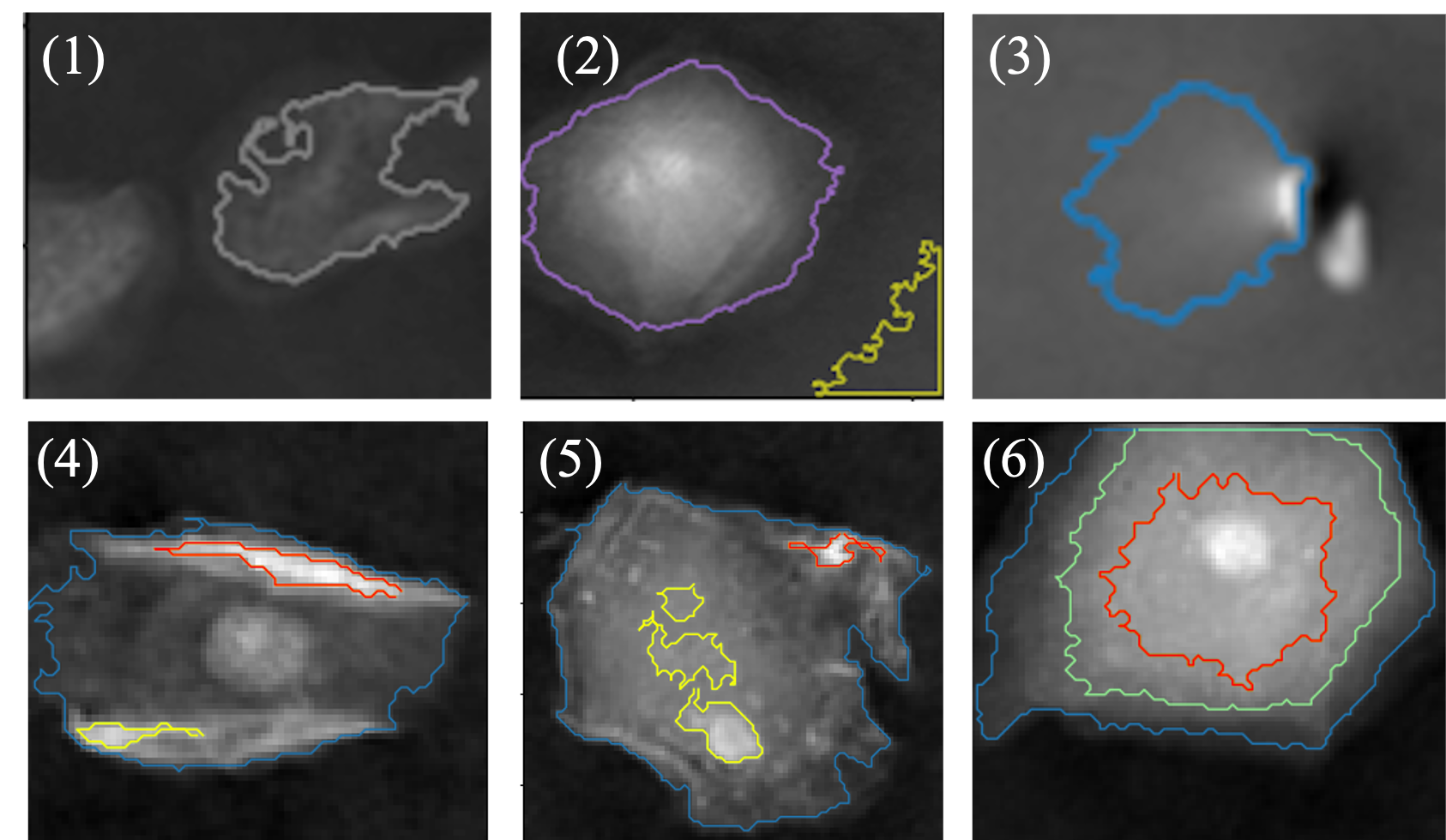}}
\caption{Examples of occuring errors. The occurring errors were categorized into the following classes, adapted from~\cite{Loewke:2018aa} to include the detection of inner structures: (1) missed cell (false negative), (2) not-a-cell (false positive), (3) poor cell boundary, (4) missed internal structure, (5) not-a-nucleus: chosen nucleus red, possible nuclei yellow, (6) poor nucleus boundary.}
\label{fig-errors}
\end{figure}

The cell and nucleus detection and feature extraction method was performed for 100 samples with a total of $8.27$ million images and $23$ million cells. In total 68 features are computed for each cell (most of which are standard features from literature; only the most relevant new features were presented here). The average analysis time was $\SI{0.113}{\s}$ per cell or $\SI{0.3}{\s}$ per image. To quantify the cell detection accuracy, the 3 images with the most detected cells were checked manually on a cell-by-cell basis for each of the samples. Similarly, the nucleus segmentation was checked for 15 random cells of each sample. The results are shown in table~\ref{tab1}, examples of typical errors are shown in figure~\ref{fig-errors}. The most important goal of the cell detection was that no relevant cells are missed, even though the appearance of cells is unknown, therefore error (1) is the most important, the results with $1.85\%$ are quite good. Most of the cells which are missed are small cells, which are discarded through step 3 of the segmentation and are not relevant for a diagnosis. Not-a-cell errors (2) and poor cell boundaries (3) occur when there is debris or slime causing very noisy images, reflections and phase wraps. These errors are more common, however, they can be detected through the computed features and dealt with in post-processing. In the internal detection also only very few structures are missed and the algorithm chooses the correct nucleus in most cases. The most common error is a poor nucleus boundary, error type (6), which is caused through cells which are badly focused or have optical density distributions which are unusually high. This again can be dealt with in post-processing for example by applying active contour methods.
\begin{table}[htbp]
\caption{Segmentation Results for 100 patient samples}
\begin{center}
\begin{tabular}{|c|c|c||c|c|c|}
\hline
\multicolumn{3}{|c||}{\textbf{Cell Detection Errors}} &\multicolumn{3}{c|}{\textbf{Internal and Nucleus Errors}} \\
\multicolumn{3}{|c||}{\textbf{for $4059$ cells}} &\multicolumn{3}{c|}{\textbf{for $1500$ cells}} \\
\textbf{(1)}& \textbf{(2)}& \textbf{(3)} & \textbf{(4)} & \textbf{(5)} & \textbf{(6)} \\
\hline
 $75$ & $241$ & $434$ & $35$ & $71$ & $199$\\
\hline
$1.85\%$ & $5.94\%$ & $10.69\%$ & $2.33\%$ & $4.73\%$ & $13.27\%$  \\
\hline
\end{tabular}
\label{tab1}
\end{center}
\end{table}

\section{Conclusion}
\label{sec:conclusion}
We presented an unsupervised multistage method for quantitative phase microscopy, which includes an automatic threshold choice based on the statistics of the detected phase values, several plausibility checks and the detection of relevant inner structures, especially the cell nucleus in the unstained cell. Additionally, we introduce new morphological features that exploit the structural  and optical density information of cells, that may eventually even touch genetic abnormalities and be of diagnostic use. We showed that the segmentation provides consistently good results over many experiments on patient samples in a reasonable per-cell analysis time.

\section*{Compliance with Ethical Standards}
This study was performed in line with the principles of the Declaration of Helsinki. Approval was granted by the Ethics Committee of the Technical University of Munich (Date 25.04.2022/No. 2022-156-S-KH).

\section*{Conflicts of Interest}
No funding was received for conducting this study. The authors have no relevant financial or non-financial interests to disclose.


\bibliographystyle{IEEEtran}
\bibliography{IEEEabrv, seg-bib}
\end{document}